\documentclass{INTERSPEECH2023}

\interspeechcameraready 

\usepackage{xcolor}

\title{Modular Domain Adaptation for Conformer-Based Streaming ASR}
\name{Qiujia Li, Bo Li, Dongseong Hwang, Tara N. Sainath, Pedro M. Mengibar}
\address{Google LLC}
\email{\{qiujia,boboli,dongseong,tsainath,pedro\}@google.com}

\begin{document}

\maketitle
 
\begin{abstract}
Speech data from different domains has distinct acoustic and linguistic characteristics. It is common to train a single multidomain model such as a Conformer transducer for speech recognition on a mixture of data from all domains. However, changing data in one domain or adding a new domain would require the multidomain model to be retrained. To this end, we propose a framework called \emph{modular domain adaptation} (MDA) that enables a single model to process multidomain data while keeping all parameters domain-specific, \emph{i.e.}, each parameter is only trained by data from one domain. On a streaming Conformer transducer trained only on video caption data, experimental results show that an MDA-based model can reach similar performance as the multidomain model on other domains such as voice search and dictation by adding per-domain adapters and per-domain feed-forward networks in the Conformer encoder.
\end{abstract}
\section{Introduction}
Automatic speech recognition (ASR) systems have been used across various applications, such as video captioning~\cite{Liao2013LargeSD}, dictation~\cite{Li2015LSTMTA}, voice search~\cite{Wang2008AnIT,Shan2017AttentionBasedES}, voice assistant~\cite{Kepuska2018NextgenerationOV} and telephony~\cite{Hain2000THECM,Xiong2017TheM2,Tuske2021OnTL}. Each domain has its own acoustic and linguistic characteristics.
For example, video captioning data has diverse acoustic environments and speaking styles; a dictation utterance is likely to have little background noise; voice search data tends to have short queries with named entities.

The performance of ASR models deteriorates significantly when the model is trained on a particular domain but evaluated on another~\cite{Bell2020AdaptationAF}. Given sufficient model capacity and training data, it is desirable to build a single ASR model to serve all application domains. To ensure the model performs well for all domains, one simple and effective approach is to mix all data during training to obtain a multidomain model~\cite{Narayanan2018TowardDS,Chan2021SpeechStewSM}. Since all model parameters are shared across all domains, the multidomain approach has some shortcomings. First, the entire model needs to be retrained if the training data from a certain domain changes or a new domain is added. Second, finding the right balance to mix data from various domains is nontrivial. On the other end of the spectrum, it is also possible to build an ASR model for each domain. This means all parameters are domain-specific where each parameter is trained on data from a single domain. However, this is very expensive for training and serving as multiple models need to be maintained. Therefore, it is ideal to build a single model that serves all domains while keeping all model parameters domain-specific. 

\begin{figure}
    \centering
    \includegraphics[width=0.9\linewidth]{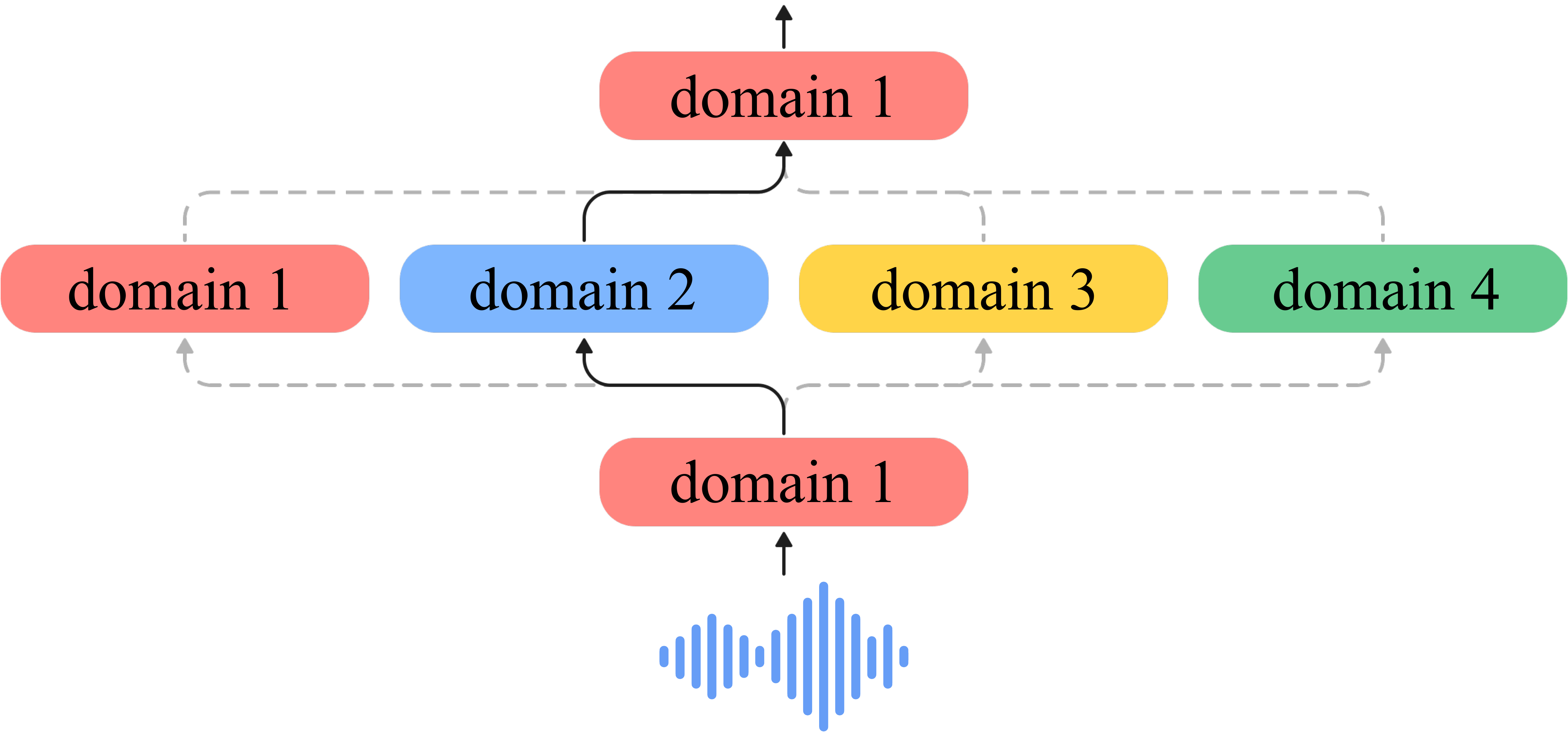}
    \vspace{-1em}
    \caption{Modular domain adaptation (MDA). Different colors indicate parameters trained on different domains. In this example, speech from domain 2 passes through the backbone trained on domain 1 except when per-domain parameters are available.}
    \label{fig:mda}
    \vspace{-2em}
\end{figure}

The contribution of this work is as follows. First, we proposed a framework called \emph{modular domain adaptation} (MDA) that can process multidomain data with all parameters being domain-specific. As illustrated in Fig.~\ref{fig:mda}, the backbone model is trained on a particular domain (in red), and all traffic passes through the common backbone model except certain parts of the model where the traffic is split and routed through the per-domain parameters. Second, we identified the most effective components of the Conformer backbone model to integrate per-domain parameters for parameter efficiency. Also, two types of per-domain parameters are explored, \emph{i.e.}, per-domain components that replace the existing components in the backbone model and per-domain adapters that add lightweight adapter modules to modify intermediate representations. Third, the final recipe was validated on three different domains with a large amount of data and minimum word error rate (MWER) training~\cite{Prabhavalkar2017MinimumWE}. The MDA-based model achieved similar performance as the multidomain model across all domains with 0.2--0.4\% absolute degradation in word error rates (WERs). The number of per-domain parameters is 22\% of the backbone model.

Although domain adaptation has been widely studied for ASR~\cite{Bell2020AdaptationAF} with various paradigms including input feature adaptation~\cite{Gales1998MaximumLL,Fainberg2017FactorisedRF,Sainath2020ASO}, model-based adaptation~\cite{Li2017LargeScaleDA,Asami2017DomainAO,Manohar2018ATL,Samarakoon2018DomainAO,Sim2018DomainAU} and multi-task learning~\cite{Denisov2018UnsupervisedDA,Meng2018AdversarialTL}, our work pays particular attention to modularity. Unlike general adaptation techniques, MDA satisfies the constraint that all model parameters are domain-specific. This brings many advantages of modularity~\cite{pfeiffer2023modular}: similar functions of the ASR model are encoded with the same module while allocating distinct functions to per-domain parameters; per-domain parameters can be constructed separately and updated locally; parameter efficiency is much higher than finetuning the entire model or have multiple single-domain models.

In this paper, two methods under the framework of MDA are described in Sec.~\ref{sec:methods}. Extensive experimental studies are then conducted in Sec.~\ref{sec:setup}~\&~\ref{sec:exp}, which offer insights into the Conformer transducers, especially in terms of modularity. Conclusions are drawn in Sec.~\ref{sec:conclusions}.
\section{Methods}
\label{sec:methods}
In this section, two types of methods for MDA are introduced. Both approaches enable the inputs to be transformed differently based on their corresponding domains. Per-domain components replace a particular part of the backbone model with separate parameters for each domain. Per-domain adapters are additional domain-specific modules added to the backbone model.
\begin{figure}[t]
    \centering
    \includegraphics[width=0.7\linewidth]{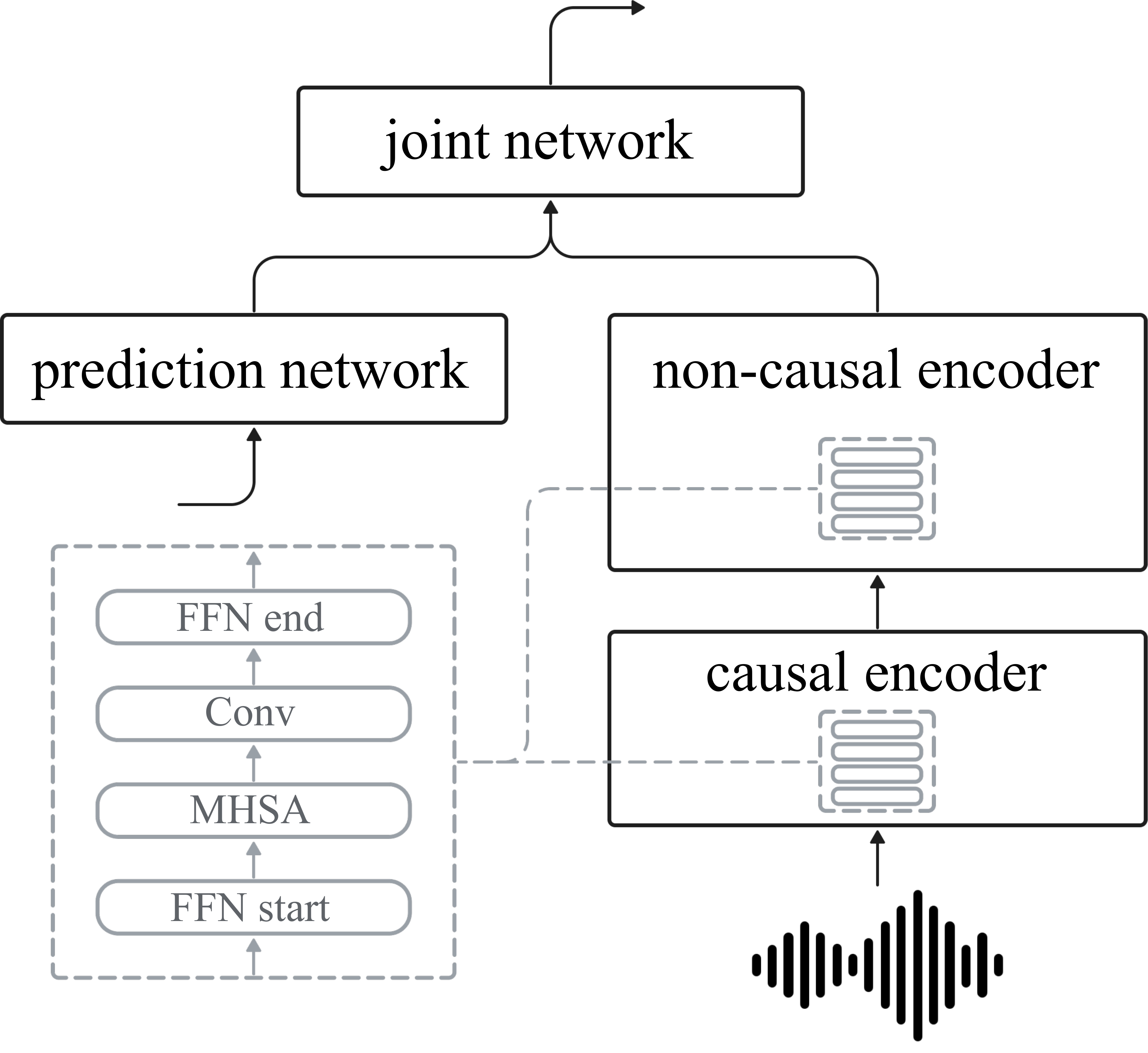}
    \caption{Conformer transducer with cascaded encoders.}
    \label{fig:conformer}
    \vspace{-1em}
\end{figure}

\subsection{Per-Domain Components}
As illustrated in Fig.~\ref{fig:conformer}, a Conformer transducer~\cite{Gulati2020ConformerCT} has an encoder with a stack of Conformer blocks and a decoder containing a prediction network and a joint network. A Conformer block contains a sequence of four modules - a feed-forward network (FFN), a multi-head self-attention (MHSA) module, a convolution module (Conv), and another feed-forward network. Each module in the Conformer block has a residual connection, which is omitted in the figure for simplicity. To enable processing streaming audio, a limited amount of future context is available for MHSA and Conv modules~\cite{Li2020ABA}.

To perform MDA, some components of the Conformer transducer can be customized for each domain, \emph{i.e.}, same architecture but with different parameters for each domain. As shown in Fig.~\ref{fig:perdomain}, the input examples are routed through different components corresponding to their domains in different colors. The granularity of a ``component'' can vary greatly. A component can be a module in a Conformer block, an entire Conformer block, or even the entire encoder. At one extreme, it is equivalent to having a model for each domain if all components of the model are separate for each domain.
Since some components will have a greater impact on the model quality than others, the effect of per-domain Conformer components will be examined in Sec.~\ref{sec:exp}.

\subsection{Per-Domain Adapters}
\begin{figure}[t]
    \centering
    \begin{subfigure}{0.32\linewidth}
        \centering
        \includegraphics[height=7em]{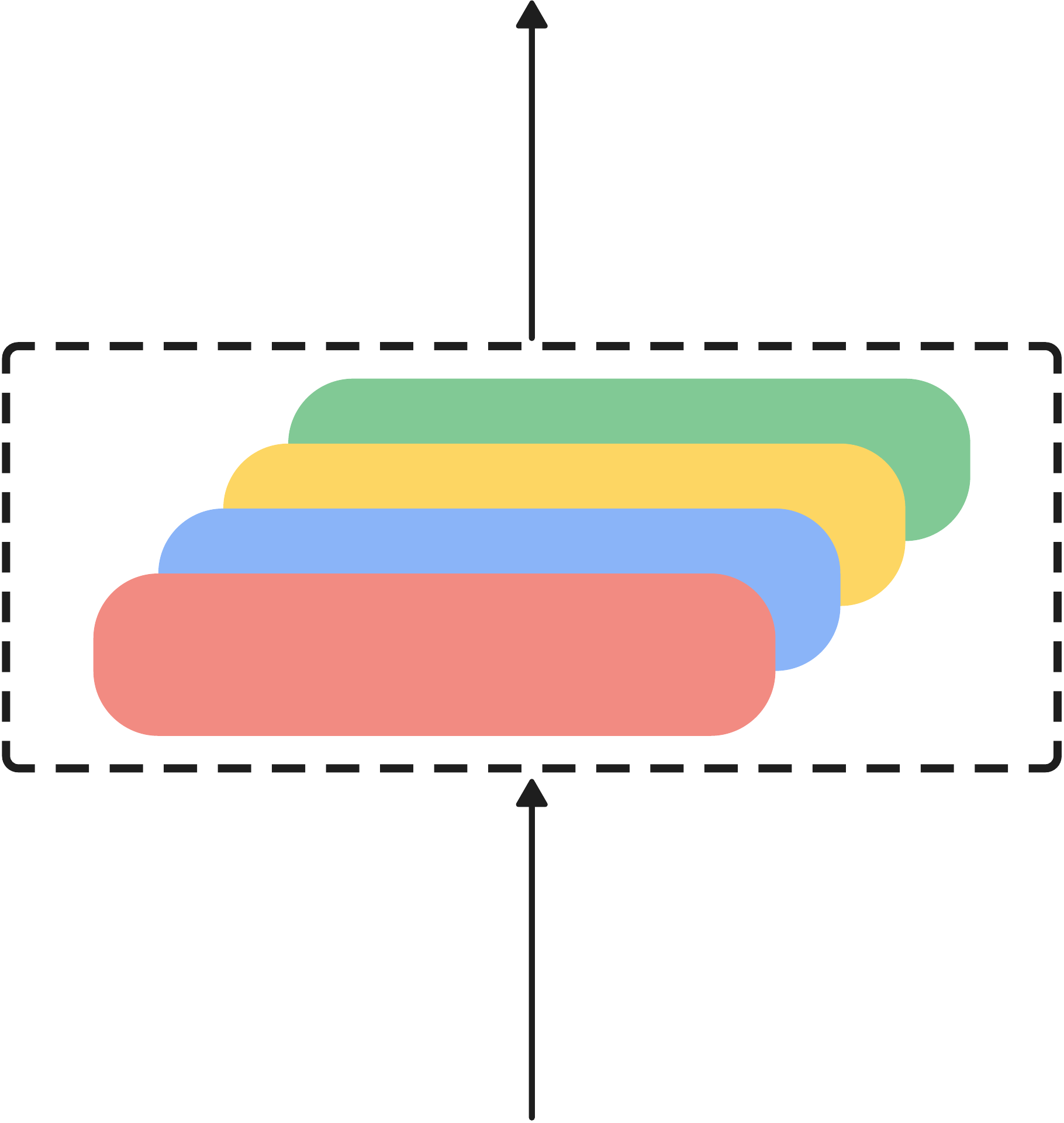}
        \captionsetup{justification=centering}
        \caption{Per-domain\\components.}
        \label{fig:perdomain}
    \end{subfigure}
    \begin{subfigure}{0.32\linewidth}
        \centering
        \includegraphics[height=7em]{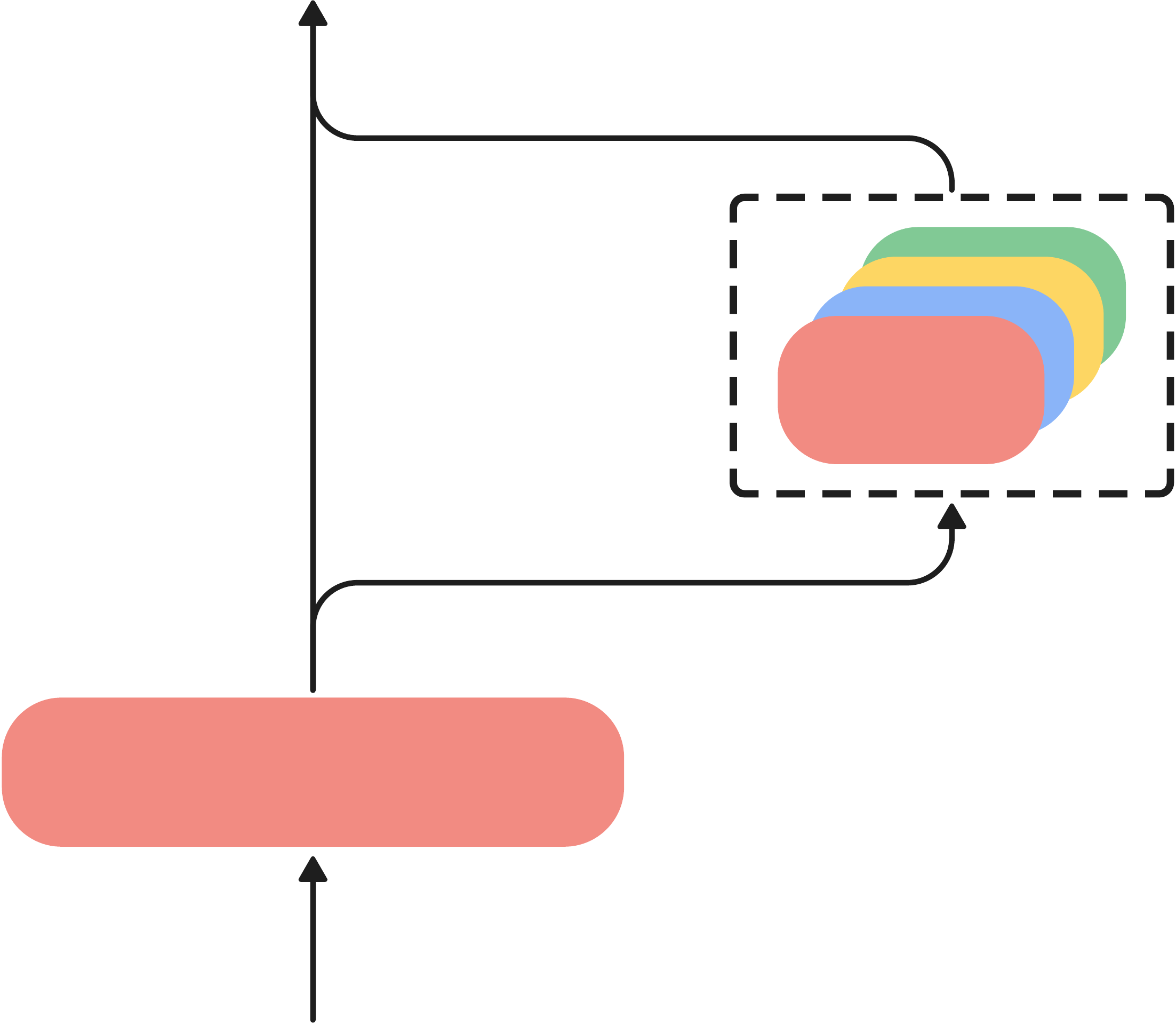}
        \captionsetup{justification=centering}
        \caption{Per-domain\\sequential adapters.}
        \label{fig:seqadapter}
    \end{subfigure}
    \begin{subfigure}{0.32\linewidth}
        \centering
        \includegraphics[height=7em]{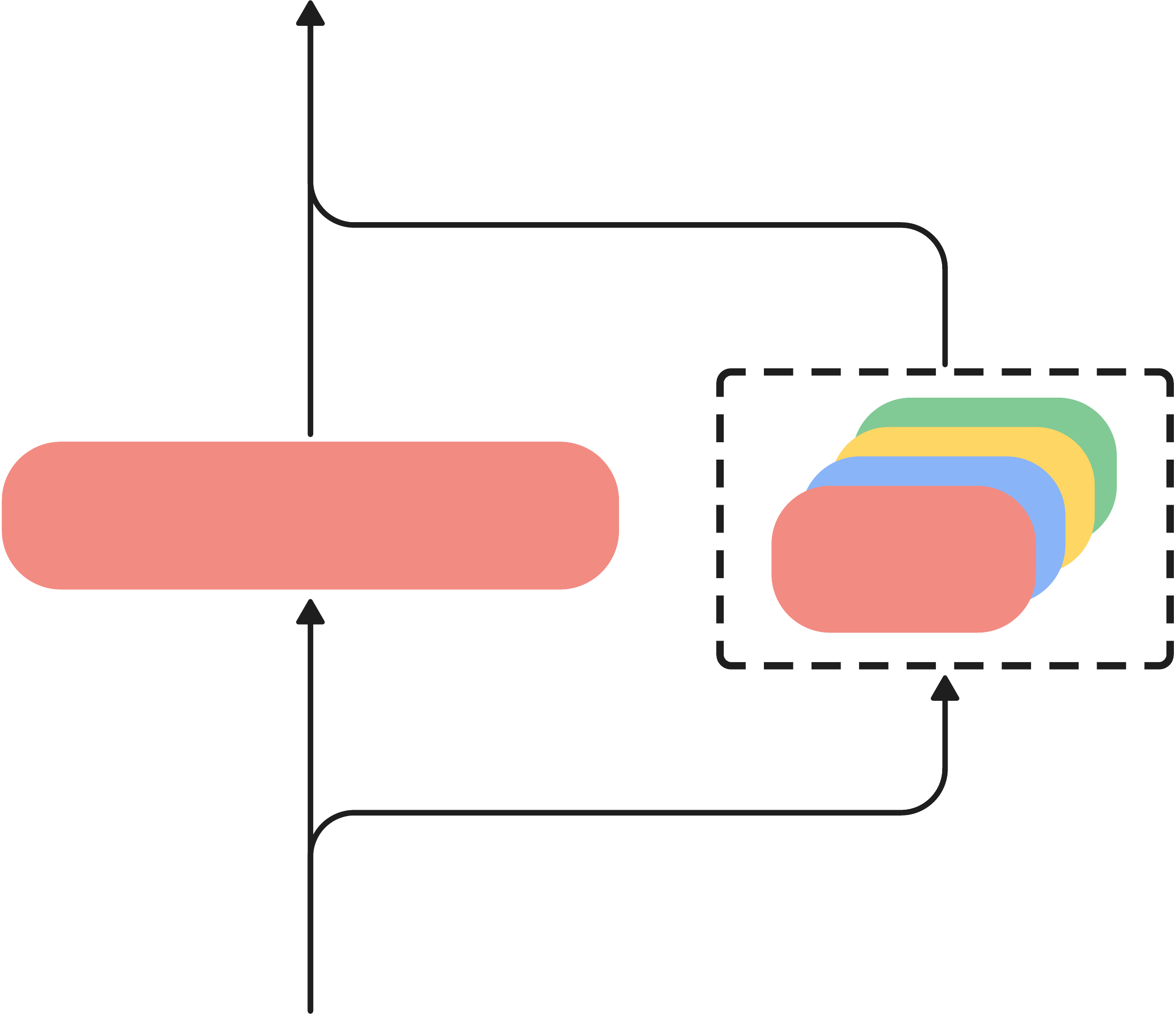}
        \captionsetup{justification=centering}
        \caption{Per-domain\\parallel adapters.}
        \label{fig:paradapter}
    \end{subfigure}
    \caption{Various modular domain adaptation (MDA) methods.}
    \vspace{-1em}
\end{figure}

Another approach to MDA is to have per-domain adapters in the Conformer model. Adapters are commonly used as efficient modules to adapt large foundation models for particular tasks without finetuning all model parameters~\cite{Houlsby2019ParameterEfficientTL,Pfeiffer2020AdapterFusionNT,He2021TowardsAU,Li2023EfficientDA}. For MDA, per-domain adapters can be used to augment the hidden representations of the backbone model accordingly. For each domain, the adapter first uses a weight matrix $\mathbf{W}_{\text{down}}\in\mathbb{R}^{d\times b}$ to project the input representation $\mathbf{h}\in\mathbb{R}^{1\times d}$ to a lower-dimensional space with bottleneck dimension $b$, followed by a nonlinear activation function $f(\cdot)$, and then projects up to the original dimension using $\mathbf{W}_{\text{up}}\in\mathbb{R}^{b\times d}$. A residual connection is also applied around the adapter:
\begin{equation}
    \mathbf{h}\leftarrow \mathbf{h} + f(\mathbf{h}\mathbf{W}_{\text{down}})\mathbf{W}_{\text{up}}.
    \vspace{-0.25em}
\end{equation}
There are different ways to configure adapters for the backbone model. The per-domain adapters can be either inserted in between two adjacent modules of the model, \emph{i.e.}, sequential adapters as shown in Fig.~\ref{fig:seqadapter}, or positioned in parallel to some components of the backbone model, \emph{i.e.}, parallel adapters as shown in Fig.~\ref{fig:paradapter}. Experimental results for transfer learning have shown parallel adapters outperform sequential ones and modifying FFN representations is more effective than other modules in Transformer models~\cite{He2021TowardsAU}. In Sec.~\ref{sec:exp}, experiments will be carried out on Conformers for MDA.

The aforementioned two approaches have their respective advantages. Per-domain components do not increase the computation cost because certain components of the backbone model are simply replaced. However, per-domain adapters have a small increase in computation cost as extra parameters are inserted into the backbone model. On the other hand, per-domain components generally need to store a much larger number of domain-specific parameters than per-domain adapters as a Conformer component is generally much larger than an adapter. 

\section{Experimental Setup}
\label{sec:setup}
\subsection{Data}
The training set contains three domains: YouTube (YT), voice search (VS) and dictation (DT). In total, there are 755M utterances or 1.1M hours. For YT, the training set has around 220M utterances or 600k hours, and the test set has 20k utterances or 380 hours. For VS, the training set has 520M utterances or 490k hours, and the test set has 9.4k utterances or 12 hours. For DT, the training set has 10M utterances or 23k hours, and the test set has 17k utterances or 39 hours.
Utterances from all domains were anonymized and hand-transcribed, except for the YT training set where the transcriptions were obtained in a semi-supervised fashion~\cite{Zhang2021BigSSLET}. Multi-condition training~\cite{Kim2017GenerationOL}, random data down-sampling to 8~kHz~\cite{Li2012ImprovingWS} and SpecAugment~\cite{Park2019SpecAugmentAS} were used during training to increase data diversity. 128-dimensional filterbank features were used as model inputs.

\subsection{Model}
The backbone model is a Conformer transducer model~\cite{Gulati2020ConformerCT}. The encoder is based on the cascaded structure~\cite{Narayanan2020CascadedEF,Sainath2022ImprovingTL} which consists of 7 causal Conformer blocks without future context followed by 10 non-causal Conformer blocks that use 900 ms of future acoustic context. The first two causal blocks do not have MHSA modules. The model dimensions are 512 and 640 for causal and non-causal encoders. For all Conformer blocks, MHSA modules have 8 attention heads; Conv modules have a convolution kernel size of 15; and the dimension of the FFN module is four times the model dimension. The causal and non-causal encoders have 47M and 99M parameters respectively. The outputs of both encoders are projected to 384 dimensions. There are two separate hybrid autoregressive transducer (HAT) decoders~\cite{Variani2020HybridAT} for causal and non-causal encoders. The output vocabulary has 4096 wordpieces~\cite{Schuster2012JapaneseAK}. The joint network that combines features from the encoder and the embedding prediction network~\cite{Botros2021TiedR} has 640 units. Each decoder has 9.5M parameters. The total number of parameters of the entire model is 165M. During training, two encoders are selected randomly with the same probability and FastEmit loss~\cite{Yu2020FastEmitLS} is applied with a scale of 0.005. All models were trained using the Lingvo toolkit~\cite{Shen2019LingvoAM} for 300k steps on $8\times8$ tensor processing units (TPUs) with a batch size of 4096.
\section{Experimental Results}
\label{sec:exp}
In this section, various MDA methods will be explored and compared with the multidomain baseline.
Note that all WERs reported are from the non-causal decoder without using the end pointer (EP)~\cite{Li2020TowardsFA} unless specified otherwise.
\subsection{Baseline Conformer Model}
The multidomain (MD) model is trained on a mixture of data from all five domains. Only for the MD baseline model, the domain ID encoded by a 16-dimensional one-hot vector is appended to the acoustic features as input to the model.
\begin{table}[ht]
    \centering
    \vspace{-0.25em}
    \caption{WERs (\%) of the multidomain and single-domain models. YT-only model will be used as the backbone model.}
    \vspace{-0.5em}
    \label{tab:baseline}
    \begin{tabular}{lrrr}
        \toprule
         & YT & VS & DT \\
        \midrule
        multidomain (MD)   & 14.0 &  4.7 &  3.4\\
        \midrule
        YT-only            & \textbf{13.9} & 25.1 &  8.3\\
        VS-only            & 43.0 &  \textbf{4.5} & 14.7\\
        DT-only           & 37.2 &  7.4 &  \textbf{3.7}\\
        \bottomrule
    \end{tabular}
    \vspace{-0.5em}
\end{table}

As shown in Table~\ref{tab:baseline}, the single-domain model performs well on in-domain testsets but struggles on out-of-domain testsets. The performance of the MD model is about the same as the single-domain models on their respective domains for YT and VS as they both have a large amount of training data. The MD model has a lower WER than the DT-only model on the DT testset because the size of the DT training set is an order of magnitude smaller than YT or VS. Since YT is considered to be the most diverse domain containing speech in a wide range of forms and acoustic conditions, the YT-only model is used as the backbone model in the rest of the paper. Note that the parameters of the backbone model will not be updated and per-domain parameters will be initialized randomly from scratch.

\subsection{Per-Domain Components}
\label{sec:domaincomponent}
In this section, various components in the backbone YT-only model are replaced by per-domain ones. For simplicity, only results on the VS testset are reported and the observations on the DT domain are similar.

\begin{figure}[ht]
    \centering
    \vspace{-1em}
    \begin{subfigure}[t]{0.4\linewidth}
        \centering
        \includegraphics[width=0.95\linewidth]{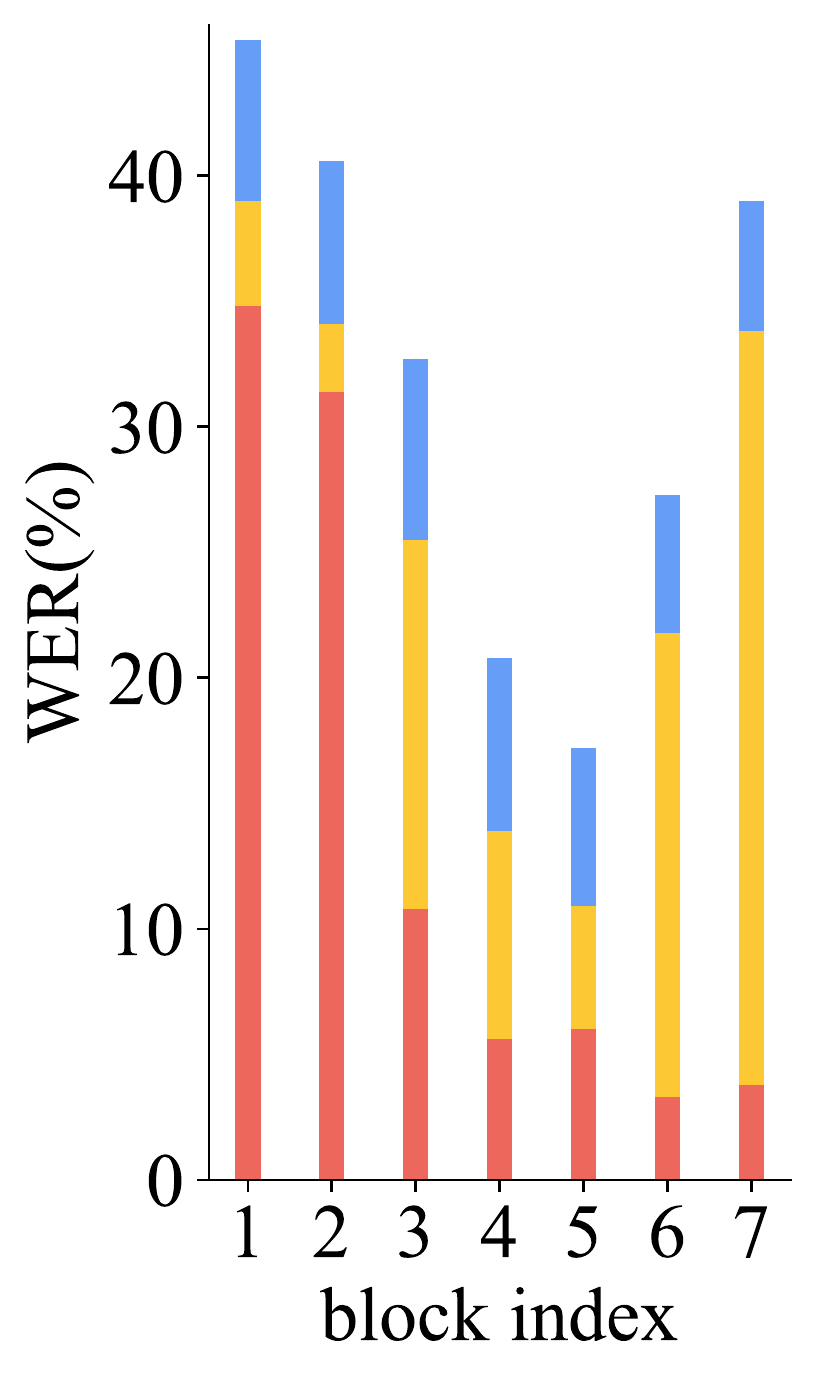}
        \caption{Causal encoder.}
        \label{fig:perdomaincausal}
    \end{subfigure}
    ~
    \begin{subfigure}[t]{0.5\linewidth}
        \centering
        \includegraphics[width=\linewidth]{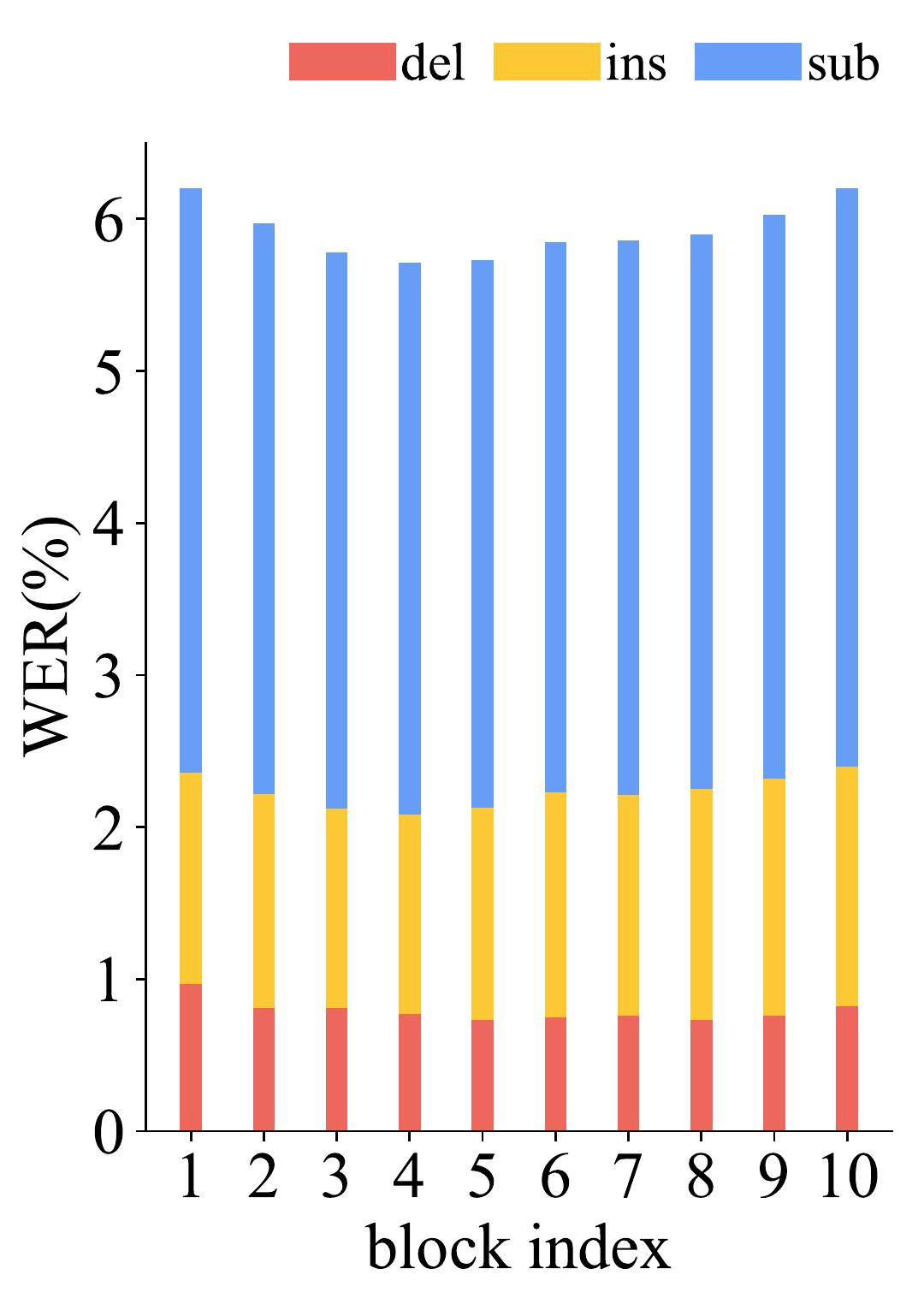}
        \caption{Non-causal encoder.}
        \label{fig:perdomainnoncausal}
    \end{subfigure}
    \vspace{-0.5em}
    \caption{WERs on VS testset using per-domain Conformer blocks. Note that the scales of the y-axis are different. A causal / non-causal Conformer block has 6M / 10M parameters.}
    \label{fig:perdomainconformer}
    \vspace{-0.5em}
\end{figure}

First, per-domain Conformer encoder blocks are examined. In Fig.~\ref{fig:perdomainconformer}, each bar shows the WER of the model where a particular Conformer block is initialized from scratch and trained on VS while the rest of the backbone model is frozen. Fig.~\ref{fig:perdomaincausal} shows that having per-domain Conformer blocks in the causal encoder is not effective as the WER is far from the multidomain baseline (4.7\%). The trend line shows an interesting behavior where earlier blocks result in high deletion error rates but later blocks cause high insertion error rates. This may be attributed to the domain characteristics. The backbone YT-only model is trained to transcribe not only the primary speakers, but also speakers in the background. For VS, only the speech from the primary speaker is supposed to be recognized. Additionally, the average duration of a YT training utterance is 3--4 times greater than VS. We hypothesize that the earlier blocks of the encoder learn to suppress or enhance background signals. By training the earlier blocks on VS, the model tends to drop the acoustic content too aggressively whereas the later blocks cannot prevent the model to transcribe all speech content. However, the picture changes greatly for per-domain Conformer blocks in the non-causal encoder. As shown in Fig.~\ref{fig:perdomainnoncausal}, with the help of future context, the WERs drop to around 6\% for all non-causal Conformer blocks. Further experiments show that having two per-domain Conformer blocks yields marginal improvement over a single one.

Instead of using per-domain Conformer blocks, it is also possible to have per-domain Conformer modules across multiple Conformer blocks in the encoder. Table~\ref{tab:perdomain_enc} shows the WERs when using different domain-specific modules, which are initialized from scratch and then trained on VS while the rest of the backbone YT-only model remains fixed. For each Conformer module, two WER results are presented where the ``NC'' column means only the non-causal encoder has per-domain modules and ``C+NC'' means both encoders have per-domain modules. Three conclusions can be drawn from the results in Table~\ref{tab:perdomain_enc}. First, by comparing the last row, ``all params'', with the MD baseline, most modeling power of the model resides in the encoders. Second, by comparing different Conformer modules in the ``C+NC'' column, the majority of the modeling capability is within the FFN start or FFN end module. Third, by comparing the ``NC'' and ``C+NC'' columns, non-causal blocks hold much more significance than causal blocks in terms of the recognition quality on the target domain. Moreover, compared with WERs shown in Fig.~\ref{fig:perdomainconformer}, having per-domain modules across all blocks is more effective than having a per-domain Conformer block. 
\begin{table}[ht]
    \centering
    \vspace{-0.25em}
    \caption{WERs on VS testset using different per-domain encoder components. ``C / NC'' mean ``causal / non-causal'' encoders.}
    \vspace{-0.5em}
    \label{tab:perdomain_enc}
    \begin{tabular}{lcrrr}
        \toprule
        & \# trainable params & \multicolumn{2}{c}{WER (\%)}\\
        \cmidrule{3-4}
        & C / NC (M) & NC & C+NC \\
        \midrule
        \textit{MD baseline} & & \multicolumn{2}{c}{4.7}\\
        \midrule
        FFN start  & 16.8 / 32.8 & \textbf{4.9} & \textbf{4.7}\\
        MHSA       &  \,\,\,4.2 / 20.5 & 5.5 & 5.3\\
        Conv       &  \,\,\,5.0 / 12.4 & 5.4 & 5.0\\
        FFN end    & 16.8 / 32.8 & 5.0 & \textbf{4.7}\\
        \midrule
        all params & 46.8 / 99.1 & 4.8 & 4.6\\
        \bottomrule
    \end{tabular}
    \vspace{-0.5em}
\end{table}

In Table~\ref{tab:perdomain_dec}, per-domain decoder components are also tested. Having a per-domain prediction network is not useful as it has a similar WER as the backbone YT-only model (25.1\%). This is expected because the prediction network only provides the token embeddings of the two previous tokens for the joint network~\cite{Botros2021TiedR}. The joint network plays the most important role within the decoder as it transforms the encoder features, but the WER of the per-domain joint network is far from the MD baseline. Further experiments have also been carried out to combine domain-specific Conformer modules and per-domain joint networks, the improvement over the results in Table~\ref{tab:perdomain_enc} is small, which further validates the finding that the main modeling capacity is within the encoder.
\begin{table}[ht]
    \centering
    \vspace{-0.25em}
    \caption{WERs on VS testset using different per-domain decoder components.}
    \vspace{-0.5em}
    \label{tab:perdomain_dec}
    \begin{tabular}{lcr}
        \toprule
         & \# trainable params (M) & WER (\%)\\
        \midrule
        \textit{MD baseline} & & 4.7\\
        \midrule
        prediction network & 6.1 & 27.0\\ 
        joint network & 3.3 & \textbf{9.2}\\
        \midrule
        all params & 9.3 & 9.3\\
        \bottomrule
    \end{tabular}
    \vspace{-0.5em}
\end{table}

\subsection{Per-Domain Adapters}
Using per-domain adapters is a more parameter-efficient approach for MDA as the number of additional domain-specific parameters is very small. Table~\ref{tab:adapters} shows the WERs on the VS testset for both sequential adapters as in Fig.~\ref{fig:seqadapter} and parallel adapters as in Fig.~\ref{fig:paradapter}. Since the FFN module is the most important module in the Conformer as found in Sec.~\ref{sec:domaincomponent}, both types of per-domain adapters are applied around all the FFN (start \& end) modules in the encoder. There are three findings from Table~\ref{tab:adapters}. First, by comparing sequential and parallel adapters of the same bottleneck dimension, parallel adapters consistently outperform sequential ones. This is consistent with the literature in other fields~\cite{He2021TowardsAU}. Second, a larger bottleneck dimension helps for both types of adapters. Third, similar to Table~\ref{tab:perdomain_enc}, the adapters in the non-causal encoder are very significant and the adapters in the causal encoder bring small but consistent gains. Compared with the per-domain Conformer blocks in Sec.~\ref{sec:domaincomponent}, per-domain adapters use much fewer domain-specific parameters while achieving lower WERs.
\begin{table}[ht]
    \centering
    \vspace{-0.25em}
    \caption{WERs on VS testset using various per-domain adapters. ``C / NC'' mean ``causal / non-causal'' encoders.}
    \vspace{-0.5em}
    \label{tab:adapters}
    \begin{tabular}{lccrrr}
        \toprule
        \multirow{2}{*}{position} & \multirow{2}{*}{dim} & \# trainable params & \multicolumn{2}{c}{WER (\%)}\\
        \cmidrule{4-5}
        & & C / NC (M) & NC & C+NC \\
        \midrule
        \textit{MD baseline} & & & \multicolumn{2}{c}{4.7}\\
        \midrule
        \multirow{3}{*}{sequential} & 64 & 1.1 / 1.7 & 6.2 & 5.9\\
        & 128 & 2.1 / 3.3 & 5.9 & 5.7\\
        & 256 & 4.2 / 6.6 & 5.7 & 5.4\\
        \midrule
        \multirow{3}{*}{parallel} & 64 & 1.1 / 1.7 & 6.0 & 5.6\\
        & 128 & 2.1 / 3.3 & 5.8 & 5.4\\
        & 256 & 4.2 / 6.6 & \textbf{5.5} & \textbf{5.1}\\
        \bottomrule
    \end{tabular}
    \vspace{-0.5em}
\end{table}

\subsection{Final Recipe}
Given that FFN modules have the most impact on WERs in the target domain and domain adapters are very parameter efficient, together with the observation that adapting the non-causal encoder is more effective than the causal encoder, we can take advantage of both MDA approaches. Table~\ref{tab:combine} shows the WERs on the VS testset with and without end pointing for three different setups. Depending on the budget of per-domain parameters and the tolerance of quality degradation compared to the MD baseline, a trade-off can be determined. As our final recipe, per-domain adapters are added to the causal encoder to ensure reasonable end-pointing latency~\cite{Li2020TowardsFA,Sainath2022ImprovingTL} by the causal encoder and per-domain FFN modules are used in the non-causal encoder to improve the recognition quality of the target domains.
\begin{table}[ht]
    \centering
    \vspace{-0.75em}
    \caption{WERs on VS testset by various MDA methods with and without end pointing. PA refers to per-domain parallel adapters and FFN refers to per-domain FFN end.}
    \vspace{-0.5em}
    \label{tab:combine}
    \begin{tabular}{ccccc}
        \toprule
        \multirow{2}{*}{C} & \multirow{2}{*}{NC} & \# trainable params & \multicolumn{2}{c}{WER (\%)} \\
        & & C / NC (M) & w/o EP & w/ EP\\
        \midrule
        \multicolumn{3}{l}{\textit{MD baseline}} & 4.7 & 6.1\\
        \midrule
        PA & PA & \,\,\,4.2 / \,\,\,6.6 & 5.1 & 6.6 \\
        PA & FFN & \,\,\,4.2 / 32.8 & \textbf{4.7} & 6.4\\
        FFN & FFN & 16.8 / 32.8 & \textbf{4.7} & \textbf{6.3}\\
        \bottomrule
    \end{tabular}
    \vspace{-0.5em}
\end{table}

Finally, as shown in Table~\ref{tab:final}, the recipe is evaluated on all three domains and MWER training~\cite{Prabhavalkar2017MinimumWE} is applied afterwards. During the MWER training, all model parameters are updated for the MD baseline, but only per-domain parameters are updated for the MDA recipe. The total number of per-domain parameters is around 37M and the WER degradation compared to the MD baseline after MWER training is 0.2--0.4\% absolute.
\begin{table}[ht]
    \centering
    \vspace{-0.75em}
    \caption{WERs of the final recipe on all three domains and the effect of the MWER training. WERs on VS are with EP.}
    \vspace{-0.5em}
    \label{tab:final}
    \begin{tabular}{lcrrr}
        \toprule
        & MWER & YT & VS & DT \\
        \midrule
        \multirow{2}{*}{MD baseline}   & \ding{55} &  14.0 & 6.1 & 3.4\\
        & \ding{51} & 13.5 & 5.7 & 3.1\\
        \midrule
        \multirow{2}{*}{MDA recipe} & \ding{55} & 14.1 & 6.4 & 3.7\\
        & \ding{51} & 13.7 & 5.9 & 3.5\\
        \bottomrule
    \end{tabular}
    \vspace{-0.5em}
\end{table}

\section{Conclusions}
\label{sec:conclusions}
For Conformer models, modular domain adaptation (MDA) approaches such as per-domain feed-forward networks and per-domain adapters can achieve similar recognition quality as a multidomain model while having the benefits of modularity by using all domain-specific parameters for the entire model. In the future, we plan to explore other MDA methods to further improve the WERs and achieve better parameter efficiency.


\bibliographystyle{IEEEtran}
\bibliography{ref}

\end{document}